\documentclass[
 reprint,
groupedaddress,
 amsmath,amssymb,
 aps,
 prmaterials
]{revtex4-2}

\usepackage{graphicx}
\usepackage{dcolumn}
\usepackage{bm}
\usepackage{amsthm,amsmath,amssymb,amsfonts}%
\usepackage{tabularx}
\usepackage{caption}
\usepackage{subcaption}
\usepackage{float}
\usepackage{booktabs}
\usepackage{caption}
\usepackage[utf8]{inputenc}
\usepackage{soul}
\usepackage{listings}%
\usepackage{threeparttable}
\usepackage{textcomp}%
\usepackage{xcolor}%
\usepackage{mathrsfs}%
\usepackage[polish,english]{babel} 
\usepackage[colorlinks,citecolor=red,urlcolor=blue,bookmarks=false,hypertexnames=true]{hyperref}

\begin{document}

\preprint{APS/123-QED}

\title{Implicit size dependence of the valence electron concentration criterion in high-entropy alloys}

\author{Dennis Boakye}
\author{Chuang Deng}%
 \email{chuang.deng@umanitoba.ca}
\affiliation{%
 University of Manitoba, 66 Chancellors Cir, Winnipeg, Manitoba, R3T 2N2, Canada
}%




\date{\today}

\begin{abstract}
The valence electron concentration (VEC) is the most widely used predictor of FCC against BCC stability in high-entropy alloys (HEAs), yet it is a compositional average carrying no information about atomic size. Using the macroscopic atom model, we show that the mixing enthalpy is almost size-blind, shifting by less than 6\% even when constituent volumes differ by a factor of 2, so that size can act only on the electron count. That action equals exactly the covariance of the atomic surface $V^{2/3}$ with the valence electron count, divided by its mean. This covariance is not free. Volume and valence are strongly anti-correlated across the elements used to build HEAs, so the size-corrected count is an affine rescaling of VEC over 265 characterized alloys and improves no prediction. VEC already encodes atomic size, which explains its success and locates its failure among large, electron-rich elements. Chemistry sets the enthalpy through one switch element.
\begin{description}
\item[Usage] High-entropy alloys; Valence electron concentration; Phase stability; macroscopic atom model
\end{description}
\end{abstract}

\maketitle


\section{Introduction}
\label{sec:intro}

The concept of high-entropy alloys (HEAs), introduced independently by Yeh et al.~\cite{yeh2004nanostructured} and Cantor et al.~\cite{cantor2004microstructural} in 2004, fundamentally expanded the paradigm of alloy design by demonstrating that near-equimolar mixtures of five or more principal elements can form stable single-phase solid solutions. The Cantor alloy \cite{cantor2004microstructural}, the equimolar CoCrFeMnNi alloy, crystallizes as a single-phase FCC solid solution and exhibits exceptional fracture toughness at cryogenic temperatures~\cite{gludovatz2014fracture}, making it a benchmark composition for the HEA field. The initial idea was that the large configurational entropy of mixing would thermodynamically stabilize disordered solid solutions over competing intermetallic phases. However, subsequent works have shown that entropy alone is neither sufficient nor always necessary to ensure single-phase stability~\cite{otto2013relative,guo2011effect}.

The addition of Al to the CoCrFeMnNi base alloy system has attracted sustained interest because Al simultaneously acts as a BCC stabilizer and a B2-ordering promoter, producing a rich sequence of phase transitions as a function of Al content~\cite{kumar2018effect,hsu2019effects,he2016precipitation,wang2012effects}. Experimental observations have consistently identified three regimes consisting of a single-phase FCC region at low Al content ($x_{\mathrm{Al}} \lesssim 8$ to $11$~at.\%), a dual-phase FCC+BCC/B2 region at intermediate Al ($\sim$11 to 25~at.\%), and a BCC/B2-dominated region at high Al ($\gtrsim 25$ to 30~at.\%)~\cite{kumar2018effect,he2016precipitation,wang2012effects}. Understanding the thermodynamic driving forces behind these transitions and predicting them from measurable elemental parameters remain central challenges in the design of HEAs.

In parallel with the development of 3$d$ transition-metal (TM) HEAs, refractory HEAs have emerged as a distinct class of materials targeting ultra-high-temperature structural applications~\cite{senkov2010refractory,senkov2011mechanical,zhou2023ultra}. Senkov et al. \cite{senkov2011mechanical} demonstrated that equimolar NbMoTaW and VNbMoTaW alloys maintain single-phase BCC solid-solution stability above 1600\,$^\circ$C and exhibit high compressive yield strengths at elevated temperatures. A study by Lin et al. \cite{lin2015effect} on the effect of Al concentration in the ductile HfNbTaTiZr showed a BCC phase with increasing yield strength at all concentrations. A comprehensive review by Miracle and Senkov~\cite{miracle2017critical} established that RHEAs differ fundamentally from 3$d$ HEAs in their atomic size distribution, electronic structure, and melting temperatures. This raises the question of whether the same thermodynamic criteria apply to both families.

Several parametric criteria have been proposed for predicting solid-solution formation. Zhang et al.~\cite{zhang2008solid} first identified the importance of the atomic size mismatch $\delta$ and the mixing enthalpy $\Delta H$ in determining whether solid-solution, intermetallic, or amorphous phases form in as-cast HEAs. Building on this, Yang and Zhang~\cite{yang2012prediction} introduced the $\Omega$ parameter to quantify the competition between entropic stabilization and enthalpic ordering. They proposed that solid-solution formation requires $\Omega \geq 1.1$ and $\delta \leq 6.6\%$. The valence electron concentration (VEC) criterion, established by Guo et al.~\cite{guo2011effect}, provides a complementary electronic-structure-based predictor where FCC phases are favored for VEC~$\geq 8$, while BCC phases dominate for VEC~$\leq 6.87$.

Guo and Liu~\cite{sheng2011phase} performed a landmark statistical analysis comparing equiatomic solid-solution-forming HEAs with equiatomic amorphous-phase-forming alloys. They established that solid solutions form when three conditions are simultaneously satisfied such that $\delta \leq 8.5\%$, $-22 \leq \Delta H \leq 7\,$kJ/mol, and $11 \leq \Delta S \leq $\,19.5\,J(K.mol). They further showed that the electronegativity difference $\Delta\chi$ has minimal discriminating power for the solid-solution versus amorphous-phase selection, while VEC plays a decisive role in determining FCC versus BCC stability within the solid-solution domain. Senkov and Miracle~\cite{senkov2016new} proposed a temperature-dependent criterion based on the parameter $k_1^{\mathrm{cr}}(T)$, which explicitly accounts for the enthalpy of formation of competing intermetallic phases. They showed that this criterion provides improved separation of solid-solution and intermetallic alloys compared to earlier $\Omega$--$\delta$ criteria. Otto et al.~\cite{otto2013relative} demonstrated through systematic substitution experiments that $\Delta S$ alone can not guarantee single-phase formation, highlighting the role of pairwise enthalpic interactions.

A distinct approach was taken by Ye et al.~\cite{ye2015design}, who proposed a single dimensionless parameter $\phi = (S_C - S_H)/|S_E|$ that incorporates not only the ideal configurational entropy and formation enthalpy but also the excessive entropy arising from dense atomic packing and atomic size misfit, computed from the Mansoori hard-sphere theory~\cite{mansoori1971equilibrium}. They showed that a critical value $\phi_c \approx 20$ separates single-phase solid solutions from multi-phase structures, providing a physically motivated design parameter that unifies the effects of composition, enthalpy, and atomic packing into a single criterion.

The macroscopic atom model (MAM), rooted in Miedema's semi-empirical theory of alloy formation~\cite{miedema1980cohesion,boer1988cohesion}, provides a physically grounded framework for computing pairwise interaction enthalpies from tabulated elemental parameters. Unlike CALPHAD approaches that require fitted binary interaction parameters from phase-diagram assessments, the MAM uses only three measurable quantities namely the adjusted electronegativity $\phi^*$, the electron density at the Wigner--Seitz boundary $n_{\mathrm{ws}}^{1/3}$, and the molar volume $V^{2/3}$ to predict the enthalpy of mixing for any binary pair~\cite{miedema1980cohesion}. The dilute-limit interaction enthalpies $\Delta H^\circ_{A \text{ in } B}$ and $\Delta H^\circ_{B \text{ in } A}$ capture the asymmetry of the interaction and can be extended to multicomponent systems through the regular or sub-regular solution formalism. The framework is particularly well-suited for RHEAs because the strong electronegativity and electron-density contrasts between early (IVB) and late (VIB) refractory elements produce well-defined enthalpic signatures. The Takeuchi--Inoue tabulation~\cite{takeuchi2005classification} provides widely used equimolar binary enthalpies derived from this framework and serves as a validation benchmark.

A feature common to all of these criteria is that they treat the mixing enthalpy $\Delta H$ and the atomic size mismatch $\delta$ as independent axes of a phase-stability map, and that the structural discriminator within the solid-solution field, VEC, is a plain compositional average containing no geometric information whatsoever. This combination is unexpected. The field has treated the absence of size information in VEC as a deficiency, to be repaired either by recalibrating its thresholds or by adding $\delta$ as an independent axis. Yang et al.~\cite{yang2020revisit} revisited the VEC rule for AlCoCrFeNi using high-throughput CALPHAD and found the original bounds too permissive, reporting instead that BCC dominates for $5.7 \leq \mathrm{VEC} \leq 7.2$ and that FCC is universal only above $\mathrm{VEC} = 8.4$; a companion high-throughput ab initio study reached a similar conclusion~\cite{yang2022revisit}. Yet no recalibration and no supplementary size axis has displaced VEC itself, which remains the single most-used structural predictor in the field and, in machine-learning studies over thousands of compositions, the most important individual feature \cite{nutor2020phase,zhang2025microstructure,boakye2026machine}. A criterion containing no geometry therefore outperforms criteria constructed around geometry, and the present work sets out to explain why.

Our route to the answer is the surface-fraction formulation of the MAM, in which the composition dependence of the enthalpy is built not from atomic fractions but from the volume-based contact fractions $s_i = c_i V_i^{2/3} / \sum_k c_k V_k^{2/3}$ that measure the surface each element presents to unlike neighbors. Atomic volume thereby enters every computed quantity through a single, physically transparent geometric weight, and the two branches of phase selection can be differentiated with respect to size on a common footing.

Four questions are examined in turn. The first is how sensitive the mixing enthalpy actually is to atomic volume, and whether that sensitivity is large enough to bear on solid-solution stability. The second is the form taken by the size correction to the electron count when the same surface weighting is applied to VEC, and which elemental quantities control its magnitude. The third is whether that correction constitutes information independent of VEC itself, which we assess against a database of 265 experimentally characterized alloys compiled from 88 sources, refitting the decision thresholds separately for each parameter so that the two are compared on equal terms. The fourth is what governs the mixing enthalpy if atomic size does not, for which the surface-fraction decomposition is resolved into its pairwise contributions across both alloy families. These questions in all determine whether atomic size constitutes an independent axis of phase selection in HEAs, or whether its influence is already contained within the criteria in current use.

\section{Computational methodology}
\label{sec:methods}

\subsection{Element set and Miedema parameters}

The elemental parameters used in this work are drawn from the standard Miedema tabulations~\cite{boer1988cohesion,miedema1980cohesion}. The full element set comprises 14 elements spanning two alloy families, namely the 3$d$ transition metals plus Al (Co, Cr, Fe, Mn, Ni, Al) and the refractory elements (Ti, Zr, Hf, V, Nb, Ta, Mo, W). Table~\ref{tab:elem_params} lists the Miedema parameters, melting temperatures, VEC values, atomic radii, and Pauling electronegativities for all 14 elements. This represents a total of 91 unique binary pairs, all of which are present in our interaction database.
\begin{table}[!ht]
\centering
\caption{Physical data for all 14 elements in this study. The values for the parameters $\phi^*$ (V), $n_{ws}\,\text{(density units)}^{1/3}$ and $V$ (cm$^3$/mol) are taken from Ref. \cite{boer1988cohesion}.}
\label{tab:elem_params}
\small
\begin{tabular}{@{}l c c c c c c c@{}}
\toprule
Element & $T_m$ (K) & VEC & $r$ (pm) & $\phi^*$ & $n_{\text{ws}}$ & $V$  & $\chi_{\mathrm{Pauling}}$ \\
\midrule
Al & 933.5  & 3  & 143 & 4.20 & 2.70 & 10.00 & 1.61 \\
Co & 1768   & 9  & 125 & 5.10 & 5.36 & 6.70  & 1.88 \\
Cr & 2180   & 6  & 128 & 4.65 & 5.18 & 7.23  & 1.66 \\
Fe & 1811   & 8  & 126 & 4.93 & 5.55 & 7.09  & 1.83 \\
Mn & 1519   & 7  & 127 & 4.45 & 4.17 & 7.35  & 1.55 \\
Ni & 1728   & 10 & 124 & 5.20 & 5.36 & 6.60  & 1.91 \\
Ti & 1941   & 4  & 147 & 3.80 & 3.51 & 10.58 & 1.54 \\
Zr & 2128   & 4  & 160 & 3.45 & 2.80 & 14.00 & 1.33 \\
Hf & 2506   & 4  & 159 & 3.60 & 3.05 & 13.45 & 1.30 \\
V  & 2183   & 5  & 134 & 4.25 & 4.41 & 8.36  & 1.63 \\
Nb & 2750   & 5  & 146 & 4.05 & 4.41 & 10.81 & 1.60 \\
Ta & 3290   & 5  & 146 & 4.05 & 4.33 & 10.81 & 1.50 \\
Mo & 2896   & 6  & 139 & 4.65 & 5.55 & 9.39  & 2.16 \\
W  & 3695   & 6  & 139 & 4.80 & 5.93 & 9.55  & 2.36 \\
\bottomrule
\end{tabular}
\end{table}

\subsection{Thermodynamic parameters}
\label{sec:thermo}

The dilute-limit interaction enthalpies $\Delta H^\circ_{i \text{ in } j}$ (the enthalpy per mole of $i$ dissolved at infinite dilution in $j$) are obtained from the Miedema model (Table S1): ~\cite{boer1988cohesion}:
\begin{equation}
	\Delta H^\circ_{i \text{ in } j} = \frac{V_i^{2/3}}{\left(n_{ws}^{-1/3}\right)_{av}}\left[-P\left(\Delta\varPhi^*_{ij}\right)^2 + Q\left(\Delta n_{ws,ij}^{1/3}\right)^2 \right]
	\label{eq:dilute}
\end{equation}
where $V_i$ is the molar volume, $n_{ws}$ is the electron density at the Wigner--Seitz cell boundary, $\Phi$ is the adjusted electron work function (Miedema electronegativity), and $Q$ and $P$ are experimentally calibrated constants with $\mathrm{Q/P=9.4\,V^2/(d.u.)^{2/3}}$. In general $\Delta H^\circ_{i \text{ in } j}\neq\Delta H^\circ_{j \text{ in } i}$, and this dilute-limit asymmetry is the physical information that a symmetric average discards.

The MAM builds the composition dependence of the mixing enthalpy not from atomic fractions but from surface (contact) fractions, which weight each element by the area its atoms present at the contact interface with unlike neighbors. For an alloy of composition $\{c_i\}$ the surface fraction of element $i$ is
\begin{equation}
	s_i = \frac{c_i\,V_i^{2/3}}{\sum_k c_k\,V_k^{2/3}},
	\label{eq:surffrac}
\end{equation}
so that large atoms ($V_i^{2/3}$ large) contribute a surface fraction that exceeds their atomic fraction. The mixing enthalpy is then the surface-fraction interpolation
\begin{equation}
	\Delta H = \sum_{i<j} c_i c_j \left[\,\hat{s}_j\,\Delta H^\circ_{i \text{ in } j} + \hat{s}_i\,\Delta H^\circ_{j \text{ in } i}\,\right],
	\quad \hat{s}_i = \frac{s_i}{s_i+s_j},
	\label{eq:dHsurf}
\end{equation}
in which each dilute-limit enthalpy is weighted by the partner's surface presence. Equation~(\ref{eq:dHsurf}) reduces to the equal-weight (symmetric) form $\Delta H = \sum_{i<j}\Omega_{ij}c_ic_j$ with $\Omega_{ij}=(\Delta H^\circ_{i \text{ in } j}+\Delta H^\circ_{j \text{ in } i})/2$ only in the equal-volume limit $V_i=V_j$, which gives $\hat{s}_i=\hat{s}_j=\tfrac12$. More generally, the common practice of assigning each pair a single composition-independent parameter $\Omega_{ij}$, as in the Takeuchi--Inoue tabulation~\cite{takeuchi2005classification} and in the criteria built upon it, amounts to freezing the surface weighting at its equimolar value rather than to discarding it. We retain the full composition dependence of Eq.~(\ref{eq:dHsurf}) throughout, and quantify the cost of both approximations in Section~\ref{sec:decoupling}. The ideal configurational entropy is $\Delta S = -R \sum c_i \ln c_i$, and the Gibbs free energy is $\Delta G = \Delta H - T\Delta S$.

The Yang--Zhang parameter $\Omega = T_m \Delta S / |\Delta H|$ (with $T_m = \sum c_i T_{m,i}$), the atomic size mismatch $\delta = 100\sqrt{\sum c_i(1-r_i/\bar{r})^2}$, and the valence electron concentration VEC~$= \sum c_i \, (\mathrm{VEC})_i$ follow standard definitions~\cite{yang2012prediction,zhang2008solid,guo2011effect}. To carry the atomic-size information that the composition-weighted VEC omits, we additionally define the surface-weighted valence-electron parameter
\begin{equation}
	\Psi = \sum_i s_i\,(\mathrm{VEC})_i,
	\label{eq:psi}
\end{equation}
which weights each element's electron count by its surface fraction rather than its atomic fraction, so that large low-VEC atoms (Al, Zr, Hf) exert proportionally more influence on the predicted FCC/BCC balance. The electronegativity difference $\Delta\chi = \sqrt{\sum c_i(\chi_i - \bar{\chi})^2}$ follows Guo and Liu~\cite{guo2011effect}. The Senkov--Miracle parameter $k_1^{\mathrm{cr}}(T) = T\Delta S/(|\Delta H| \times 1000)(1-k_2) + 1$ uses $k_2 = 0.6$~\cite{senkov2016new}. The Ye $\phi$-parameter $\phi = (S_C - S_H)/|S_E|$, where $S_H = |\Delta H|/T_m$ and $S_E$ is the Mansoori excess entropy~\cite{ye2015design,mansoori1971equilibrium}, is computed as the average of FCC ($\eta = 0.7405$) and BCC ($\eta = 0.6802$) packing fractions.

\subsection{Experimental phase database}
\label{sec:database}

To test the structural predictors against observation rather than against one another, we assembled a database of experimentally characterized alloys from the primary literature. It contains 275 entries drawn from 91 independent studies, covering 252 distinct compositions across 18 elements, all of which are present in the Miedema tabulation used here. Each entry records the composition, the processing state, the annealing temperature where applicable, the reported phase constitution, and the source (Table S2).

Two exclusions are applied before any statistical analysis. Entries produced as magnetron co-sputtered films or as-deposited coatings are removed, because such material is quenched far from equilibrium and its phase constitution is not comparable with the cast, homogenized and annealed bulk alloys against which the VEC thresholds were calibrated; this removes 10 entries. The remaining 265 entries comprise 91 single-phase BCC, 58 single-phase FCC, 58 duplex FCC$+$BCC, 52 intermetallic or multiphase, and 6 HCP-containing alloys. The binary FCC-versus-BCC discrimination task therefore uses 149 alloys and the ordinal three-class task 207.

Thresholds are never carried across parameters. Where two predictors are compared, the decision boundary for each is refitted on the same data by exhaustive search over all candidate values, since a threshold calibrated for one parameter has no meaning applied to another. Generalization is assessed by stratified five-fold cross-validated Receiver Operating Characteristic - Area Under the Curve (ROC-AUC) using logistic regression on standardized features, and the incremental value of an added parameter by a likelihood-ratio test against the nested model.

\subsection{Composition space sampling}

For the 3$d$ HEA family, we scan the Al$_x$(TM base) trajectory from 0 to 50~at.\% Al for the senary system (CoCrFeMnNi base), all five quinary subsystems (removal of one TM), all ten quaternary subsystems (three TMs), and selected non-equimolar variants (Ni-rich, Cr-lean). For the RHEA family, we scan the same Al$_x$ trajectory for 11 base systems, namely TiZrHfMoCr, TiZrNbMoCr, TiZrTaMoCr, TiZrVMoCr, TiZrVNbCr, TiVNbMoCr, ZrVNbMoCr, TiZrVNbMo, NbMoTaW, VNbMoTaW, and HfNbTaTiZr. For the full senary composition space, 15{,}000 compositions are generated by Dirichlet sampling ($\alpha_i = 1$).

\section{Results and Discussion}
\label{sec:results}

\subsection{The mixing enthalpy is size-blind}
\label{sec:decoupling}

\begin{figure*}[!ht]
	\centering
	\includegraphics[width=1\linewidth]{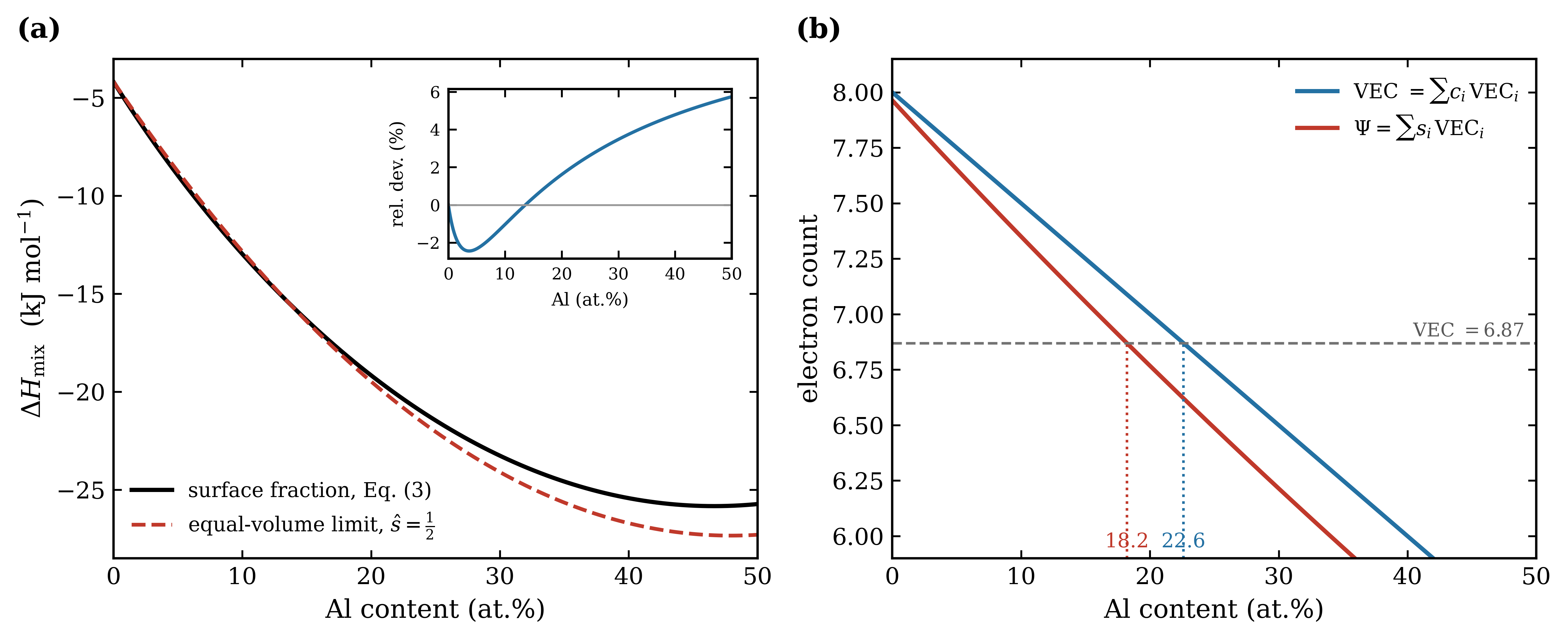}
	\caption{Atomic size acts on the electronic branch of phase selection, not the enthalpic one, for Al$_x$CoCrFeMnNi. (a) The mixing enthalpy is size-insensitive, since the surface-fraction result (solid) and its equal-volume symmetric limit (dashed) differ by less than 6\% over the whole range; the inset gives the relative deviation, which changes sign near 13\,at.\% Al. (b) The electron count is size-sensitive, so that surface-weighting the valence-electron parameter ($\Psi$, red) moves the crossing of the VEC $=6.87$ boundary from 22.6 to 18.2\,at.\% Al relative to the composition-weighted VEC (blue). Section~\ref{sec:collinear} shows that most of this apparent shift is an artifact of transferring a threshold between parameters.}
	\label{fig:decoupling}
\end{figure*}

Because the surface-fraction interpolation of Eq.~(\ref{eq:dHsurf}) is the only place atomic volume enters the enthalpy, comparing it against its equal-volume (symmetric) limit isolates exactly how size-sensitive the mixing enthalpy is. Figure~\ref{fig:decoupling}(a) makes this comparison along the Al$_x$CoCrFeMnNi trajectory. The two curves are nearly indistinguishable. The surface-fraction correction stays below 3.5\% up to 30\,at.\% Al and reaches only 5.7\% at 50\,at.\% Al, and it changes sign, so no monotonic bias is introduced. The same holds across the refractory family, where even TiZrHfMoCr, with a near two-fold spread in atomic volume, shows only a 2.2\% enthalpy shift at equimolar composition.

The reason is exact rather than coincidental, and it is not the one that inspection of the pairwise data would suggest. Subtracting the equal-weight from the surface-fraction result, the correction to any pair factorizes as
\begin{equation}
	\Delta H^{\mathrm{surf}}_{ij} - \Delta H^{\mathrm{sym}}_{ij}
	= \left(s_i - \tfrac{1}{2}\right)
	  \left(\Delta H^\circ_{j \text{ in } i} - \Delta H^\circ_{i \text{ in } j}\right),
	\label{eq:factorise}
\end{equation}
a product of two terms. The second term, the dilute-limit asymmetry, is not small. It averages 14\% across the 91 pairs and reaches 32\,kJ/mol for Al--Zr ($-208$ against $-240$\,kJ/mol), so the asymmetry that a symmetric average discards is real and substantial. What is small is the first term. Because the $2/3$ power compresses volume ratios, a two-fold ratio of molar volumes becomes only a factor $2^{2/3}=1.59$ in $V^{2/3}$, and the surface fraction is held close to one half for every pair in the database ($|s_i - \tfrac12| \le 0.123$, mean $0.048$ over all 91 pairs). The product of a small weighting deviation and a moderate asymmetry is therefore small, and the mixing enthalpy is almost size-blind. The corollary is that the approximation would degrade only for a pair combining a large volume ratio and a large dilute-limit asymmetry. The nearest approaches in our database are the Zr- and Hf-bearing pairs Ni--Zr, Co--Zr and Ni--Hf, and even these shift by no more than 4.4\%.

This has an immediate consequence for the wider literature. The Yang--Zhang, Guo--Liu and Senkov--Miracle criteria all evaluate $\Delta H$ with a single composition-independent pair parameter, which as noted in Section~\ref{sec:thermo} freezes the surface weighting at its equimolar value. Measured against the full composition-dependent treatment, this costs approximately 5\% across the Al$_x$ trajectory, and the stricter equal-weight approximation approximately 6\%. The enthalpies underlying these criteria are therefore reliable for the alloy families studied here. It also sets the logic of everything that follows. Since $\Delta H$ carries almost no size information, whatever size dependence phase selection possesses must enter through the electron count.

\subsection{The size correction to the electron count has a closed form}
\label{sec:vec}

Figure~\ref{fig:decoupling}(b) contrasts the composition-weighted VEC with the surface-weighted $\Psi$ of Eq.~(\ref{eq:psi}) along the same trajectory. Because Al is both large and low in valence (VEC $=3$), surface-weighting amplifies its influence and $\Psi$ falls faster than VEC, the gap reaching $0.31$ electrons at 50\,at.\% Al. Applied at the published VEC $=6.87$ boundary, this moves the predicted crossing from 22.6 to 18.2\,at.\% Al.

The magnitude of the displacement is not arbitrary and admits an exact closed form. Writing $v_i = V_i^{2/3}$, the difference between the two electron counts is
\begin{eqnarray}
	&&\Psi - \mathrm{VEC} \;=\; \frac{\mathrm{Cov}_c\!\left(v,\,\mathrm{VEC}\right)}{\bar{v}}, \nonumber\\
	&&\mathrm{Cov}_c(v,\mathrm{VEC}) = \sum_i c_i\left(v_i-\bar{v}\right)\left(\mathrm{VEC}_i-\overline{\mathrm{VEC}}\right),
	\label{eq:cov}
\end{eqnarray}
that is, the surface correction to the electron count is exactly the composition-weighted covariance of the atomic surface $v = V^{2/3}$ with the valence electron count, normalized by the mean atomic surface $\bar{v}$. The relevant quantity is $V^{2/3}$ and not the atomic volume itself, and the normalization is not optional, since replacing $v$ by $V$ wrongly estimates the correction by 54\% on average across the composition space. We have verified Eq.~(\ref{eq:cov}) numerically against the full calculation over $5\times10^{5}$ randomly sampled compositions spanning both alloy families, and it holds to machine precision ($\max|\text{LHS}-\text{RHS}| < 10^{-14}$).

Equation~(\ref{eq:cov}) overturns an intuition encouraged by the $\delta$-based literature. The size correction to the electron count is governed not by the size mismatch, but by whether large atoms happen also to be electron-poor. In the 3$d$ Al-HEAs the two properties are strongly anti-correlated, since Al is simultaneously the largest atom and the one of lowest valence, so the covariance is large and negative. In the refractory alloys the opposite holds. Their constituent volumes span a factor of two, giving them among the largest $\delta$ of any system studied, yet their valence electron counts are confined to the narrow band 4 to 6, so the covariance nearly vanishes and the surface correction is negligible ($-0.022$ electrons for NbMoTaW at $\delta = 2.5\%$, $-0.025$ for HfNbTaTiZr at $\delta = 4.3\%$). A large size mismatch does not by itself imply a size-biased electron count. Consistent with this, the correlation between $\delta$ and the surface correction across our systems is weak ($r = 0.54$), whereas Eq.~(\ref{eq:cov}) accounts for it exactly.

The natural reading of Fig.~\ref{fig:decoupling}(b) and Eq.~(\ref{eq:cov}) together is that $\Psi$ is the size-aware structural discriminator that VEC lacks, and that the 4.4\,at.\% displacement of the boundary is a correction the VEC criterion has been missing. That reading does not survive examination, for the reason developed in the following section.

\subsection{Volume and valence are collinear across the HEA design space}
\label{sec:collinear}

Equation~(\ref{eq:cov}) makes the size correction proportional to a covariance between two elemental properties. Whether that covariance constitutes new information depends entirely on whether the two properties vary independently across the elements from which HEAs are built. For the elements concerned, they do not.

\begin{figure*}[!htb]
	\centering
	\includegraphics[width=1\linewidth]{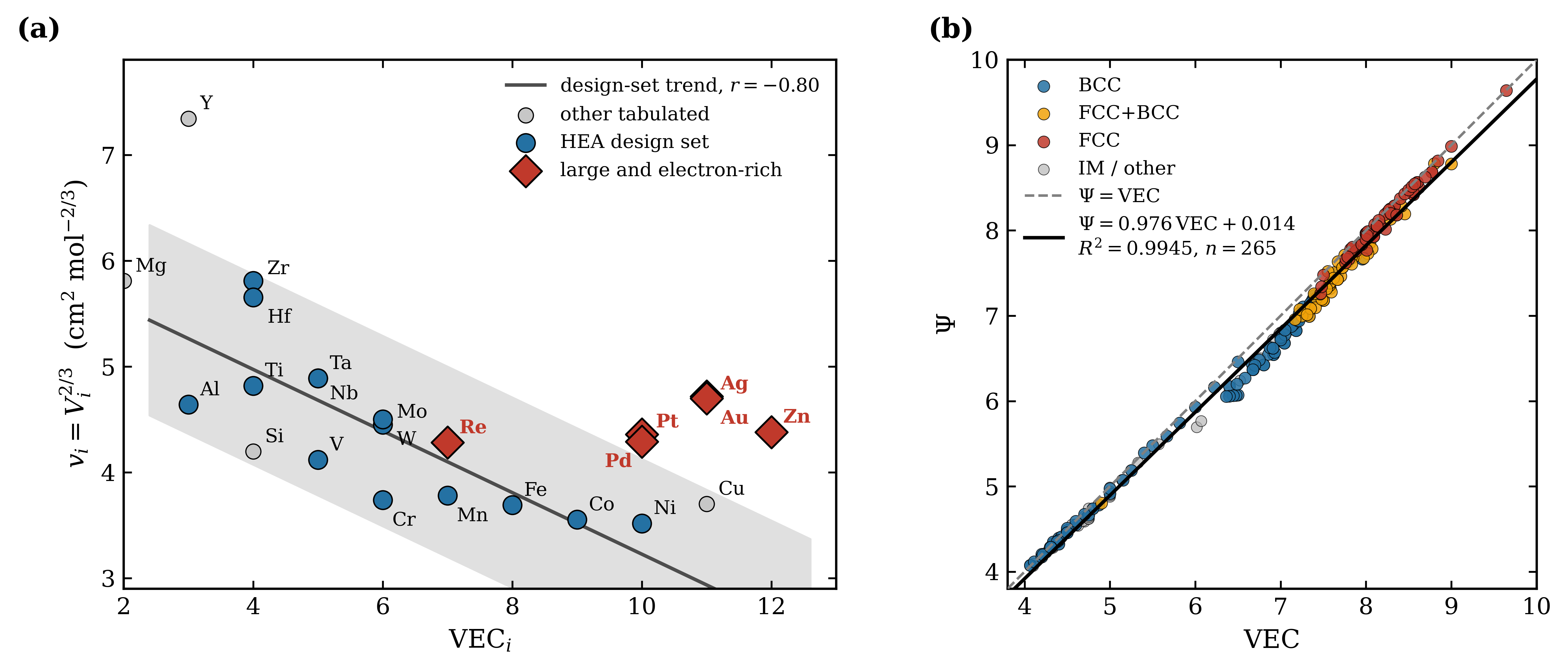}
	\caption{Volume--valence collinearity. (a) Atomic surface $v_i = V_i^{2/3}$ against VEC$_i$ for the tabulated elements. The fourteen elements of the HEA design set (blue) follow a single descending trend ($r=-0.80$; band, $\pm 2$ residual standard deviations). Elements that are simultaneously large and electron-rich (red diamonds) lie above it and break the collinearity. (b) $\Psi$ against VEC for the 265 alloys of the experimental database, colored by reported phase. The relation is affine to $R^{2}=0.9945$, so $\Psi$ carries essentially no ordering information beyond VEC.}
	\label{fig:collinear}
\end{figure*}

Figure~\ref{fig:collinear}(a) plots the atomic surface $v_i = V_i^{2/3}$ against VEC$_i$ for the tabulated elements. The fourteen elements of the present design set fall on a single descending line with $r = -0.80$, in that the Group~IVB metals and Al are large and electron-poor, the late 3$d$ metals Fe, Co and Ni are small and electron-rich, and the Group~VB and VIB refractories interpolate between them. This is not a coincidence of the particular set chosen. It reflects the systematic contraction of the metallic radius with $d$-band filling along each transition series, and it persists, though more weakly, over all 24 tabulated elements ($r = -0.56$). Because $v_i$ is, to good approximation, a decreasing linear function of VEC$_i$, the covariance in Eq.~(\ref{eq:cov}) is not free, and is instead slaved to the composition through VEC itself.

The consequence is directly visible in the experimental data. Figure~\ref{fig:collinear}(b) plots $\Psi$ against VEC for all 265 alloys of the experimental database. The points lie on a straight line,
\begin{equation}
	\Psi = 0.976\,\mathrm{VEC} + 0.014, \qquad R^{2} = 0.9945,\; n = 265,
	\label{eq:affine}
\end{equation}
with a Spearman rank correlation of $0.998$. Within this design space $\Psi$ is an affine rescaling of VEC. It preserves the ordering of alloys almost perfectly, and a parameter that preserves ordering cannot improve a classification that depends only on ordering.

\subsection{The size correction does not improve phase prediction}
\label{sec:validation}

\begin{figure*}[!ht]
	\centering
	\includegraphics[width=1\linewidth]{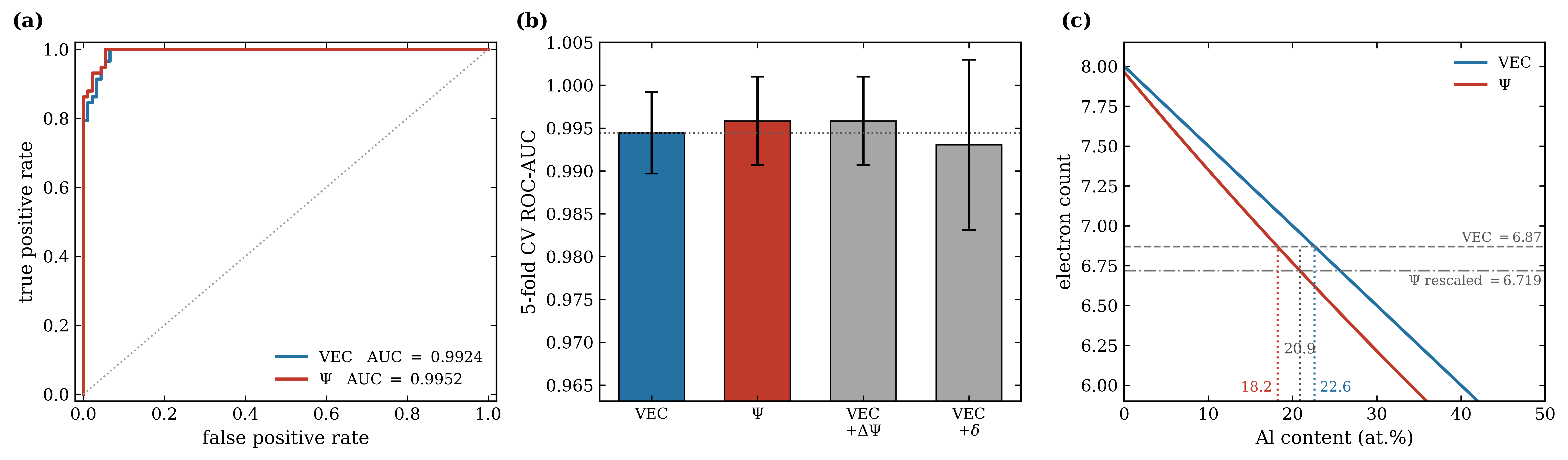}
	\caption{The size correction does not improve phase prediction. (a) Receiver-operating characteristics for FCC versus BCC discrimination over 149 single-phase alloys. (b) Stratified five-fold cross-validated ROC-AUC for VEC, $\Psi$, and VEC supplemented by the size correction $\Delta\Psi$ or by the size mismatch $\delta$; the dotted line marks the VEC baseline. (c) The Al$_x$CoCrFeMnNi crossing under the published threshold (VEC $=6.87$, dashed) and under the same threshold mapped onto $\Psi$ through Eq.~(\ref{eq:affine}) ($6.719$, dash-dotted). The displacement falls from 4.4 to 1.7\,at.\% Al once the threshold is rescaled.}
	\label{fig:validation}
\end{figure*}

Figure~\ref{fig:validation} tests this directly. Panel (a) gives receiver-operating characteristics for the discrimination of single-phase FCC from single-phase BCC across the 149 alloys for which that assignment is unambiguous. VEC achieves an area under the curve of $0.9924$ and $\Psi$ of $0.9952$, a difference of $0.003$ on a task where both are already near-perfect. Panel (b) shows the corresponding stratified five-fold cross-validated AUC, which is $0.9944 \pm 0.0048$ for VEC and $0.9958 \pm 0.0052$ for $\Psi$, with the difference an order of magnitude smaller than the fold-to-fold spread. Adding the size correction $\Delta\Psi$ to VEC as a second feature gives $0.9958 \pm 0.0052$, and adding $\delta$ gives $0.9931 \pm 0.0099$. Neither the size-corrected electron count nor the size mismatch itself changes the predictive performance of VEC by a measurable amount.

Refitting decision boundaries rather than transferring them tells the same story. On the binary task the optimal VEC threshold is $7.468$ (accuracy $0.960$) and the optimal $\Psi$ threshold $7.260$ (accuracy $0.966$), the two differing by $0.208$, which is simply the mean value of $\Psi - \mathrm{VEC}$ over the alloys near the boundary. On the ordinal three-class task the optimal thresholds are $7.217/7.877$ for VEC and $6.954/7.794$ for $\Psi$, with accuracies of $0.826$ and $0.850$; the ambiguous band between the single-phase fields is $0.660$ wide for VEC and $0.840$ for $\Psi$, so on this measure the surface-weighted parameter is the poorer discriminator.

Panel (c) resolves the boundary shift of Fig.~\ref{fig:decoupling}(b). The value $6.87$ is not a physical constant, but a threshold fitted to VEC, and applying it unchanged to a different parameter is not a meaningful comparison. Mapping it through Eq.~(\ref{eq:affine}) gives the corresponding $\Psi$ threshold of $6.719$, at which the Al$_x$CoCrFeMnNi crossing occurs at $20.9$\,at.\% Al rather than $18.2$. Of the apparent $4.4$\,at.\% displacement, $2.7$\,at.\% is therefore an artifact of transferring the threshold, and only $1.7$\,at.\% survives a fair comparison, which is well inside the scatter of reported FCC$\to$BCC transitions in this system. For comparison, in the closely related Al$_x$CoCrFeNi alloy Kao et al.~\cite{kao2009microstructure} place the duplex field at $0.45 \leq x \leq 0.88$ for as-cast material but at $0.30 \leq x \leq 1.17$ after homogenization, a processing-induced shift of the single-phase BCC onset from $18.0$ to $22.6$\,at.\% Al. The spread introduced by heat treatment alone therefore exceeds the entire disputed correction.

One methodological point deserves comment, because it bears on how parametric criteria are commonly justified. A likelihood-ratio test finds $\Delta\Psi$ significant when added to VEC in the logistic model ($p = 0.002$), yet cross-validated performance is unchanged. In-sample significance of an additional parameter within a strongly collinear feature set is not evidence of predictive value, and given the number of phase-selection parameters that have been proposed on that basis, the distinction deserves closer attention than it commonly receives.

\subsection{Why VEC works, and where it must fail}
\label{sec:whyvec}

Sections~\ref{sec:decoupling} to~\ref{sec:validation} explain a criterion whose success has been treated as empirical good fortune. VEC appears to discard atomic size, and the field has read that as a deficiency requiring either recalibrated thresholds~\cite{yang2020revisit,yang2022revisit} or a supplementary size axis. Neither diagnosis is correct. The mixing enthalpy carries almost no size information (Section~\ref{sec:decoupling}); all of it resides in the electron count, in the exact form of Eq.~(\ref{eq:cov}); and in the elements from which HEAs are made, that correction is collinear with VEC itself. VEC is not size-blind. It encodes atomic size implicitly, through the volume--valence collinearity of the elemental data, and this is why a plain compositional average outperforms criteria built explicitly around geometry.

The argument stated in this form is falsifiable, and it identifies its own failure domain, since VEC must cease to encode size wherever an alloy contains an element that departs from the master line of Fig.~\ref{fig:collinear}(a), that is, one which is simultaneously large and electron-rich. The tabulation identifies these unambiguously as Pd ($v = 4.29$, VEC $=10$), Pt ($4.36$, $10$), Ag ($4.72$, $11$), Au ($4.70$, $11$) and Zn ($4.38$, $12$), with Re ($4.28$, $7$) marginal. For such alloys the covariance in Eq.~(\ref{eq:cov}) changes sign relative to the Al-bearing systems, $\Psi$ rises above VEC rather than falling below it, and the two criteria diverge in a direction no rescaling can absorb.

None of these elements appears in our 265-alloy database, and only a single entry contains Pt. This is not a limitation of our compilation but a property of the field, in that HEA development has concentrated on the 3$d$ transition metals, Al, and the Group~IVB--VIB refractories, all of which lie on the master line. The predicted deficiency of VEC has therefore never been in a position to be observed. Noble-metal HEAs developed for catalysis, which routinely combine Pd, Pt, Ag and Au with smaller 3$d$ metals, occupy precisely the region where the collinearity breaks, and constitute the natural test bed. We propose this as the decisive experiment called for by the present analysis, since its outcome would discriminate between VEC and $\Psi$ in a way that no alloy in the current literature is able to.

\subsection{Pairwise enthalpy landscape}

\begin{figure*}[!htbp]
	\centering
	\includegraphics[width=1\linewidth]{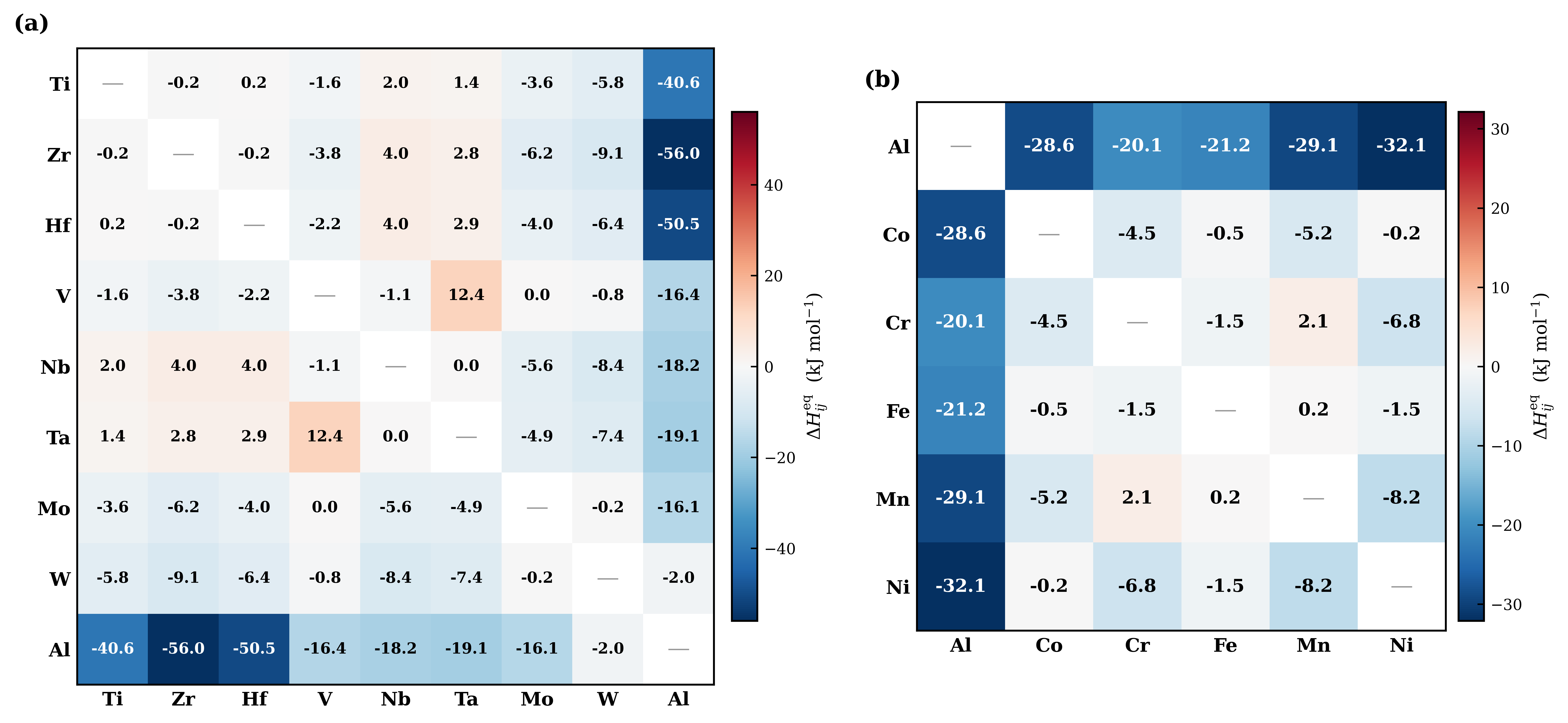}
	\caption{Equimolar binary mixing enthalpy $\Delta H_{ij}^{\mathrm{eq}} = \Omega_{ij}/4$ (kJ\,mol$^{-1}$) for (a) the refractory elements with Al and (b) the 3$d$ elements with Al. Note the order-of-magnitude contrast between the Al--TM and TM--TM blocks, and the anomalous weakness of Al--W.}
	\label{fig:heatmap}
\end{figure*}

Figures~\ref{fig:heatmap}(a) and (b) present the equimolar binary enthalpy matrices for the refractory and 3$d$+Al systems, respectively. Within the 3$d$ block, the most striking feature is the extreme magnitude contrast between Al--TM and TM--TM interactions. The Al--TM equimolar enthalpies range from $-20.1$\,kJ\,mol$^{-1}$ (Al--Cr) to $-32.1$\,kJ\,mol$^{-1}$ (Al--Ni), values that are 3 to 60 times larger than the TM--TM enthalpies, which span from $-8.2$\,kJ\,mol$^{-1}$ (Mn--Ni) to $+2.1$\,kJ\,mol$^{-1}$ (Mn--Cr). The TM--TM pairs fall into three physically distinct categories based on their Miedema electronegativity contrast where strongly negative pairs (Mn--Ni, Cr--Ni, Co--Mn) with $|\Delta\phi^*| > 0.5$~V, weakly negative pairs (Co--Cr, Fe--Cr, Fe--Ni, Co--Fe, Co--Ni) with small $|\Delta\phi^*|$ indicating near-ideal mixing, and the sole positive pair Mn--Cr ($+2.1$\,kJ\,mol$^{-1}$) arising from similar $\phi^*$ but dissimilar $n_{\mathrm{ws}}$.

The physical origin of the extreme Al--TM contrast lies in the Miedema framework itself. Al has an adjusted electronegativity of $\phi^* = 4.20$\,V and a low electron density ($n_{\mathrm{ws}} = 2.70$), while the 3$d$ TMs span $\phi^* = 4.45$ to $5.20$\,V and $n_{\mathrm{ws}} = 4.17$ to $5.55$. The large differences in both parameters produce a strong charge-transfer driving force coupled with an electron-density mismatch contribution, explaining why Al--TM interactions are an order of magnitude stronger than TM--TM interactions, where both $\phi^*$ and $n_{\mathrm{ws}}$ differences are small. The Mn--Cr positive interaction is noteworthy as it is the only pair opposing solid-solution formation, and its origin lies in the combination of similar $\phi^*$ (4.45~V for Mn against 4.65~V for Cr) but substantially different $n_{\mathrm{ws}}$ (4.17 against 5.18). This positive interaction contributes to the known tendency of Cr and Mn to drive $\sigma$-phase formation in Cr--Mn-rich HEAs~\cite{otto2013relative}.

The present TM--TM enthalpies reproduce the widely used Takeuchi--Inoue tabulation~\cite{takeuchi2005classification} closely, with $-5.2$ against $-5$ for Co--Mn, $-6.8$ against $-7$ for Cr--Ni and $-8.2$ against $-8$\,kJ\,mol$^{-1}$ for Mn--Ni, whereas the Al--TM values are larger by a factor of 1.5 to 2. We stress that this is a consistency check on the implementation and not a test of accuracy, since the Takeuchi--Inoue values are themselves computed from the Miedema model and for the liquid phase rather than for a solid solution. Two calculations sharing a model cannot validate one another. The accuracy of the enthalpies is assessed against calorimetric measurement in Section~\ref{sec:robust}, and the Al--TM divergence is traced there to the description of Al rather than to charge transfer.

The refractory pairwise landscape in Fig.~\ref{fig:heatmap}(a) differs qualitatively. All values quoted here are equimolar binary enthalpies $\Delta H^{\mathrm{eq}}_{ij} = \Omega_{ij}/4$, the same convention used for the 3$d$ block above and by Takeuchi and Inoue~\cite{takeuchi2005classification}, so that every pair in this work is reported on a single common scale. Within the refractory block the strongest interactions occur between Group~IVB and Group~VIB elements, namely Zr--Cr ($-12.4$\,kJ/mol), Hf--Cr ($-9.5$\,kJ/mol) and Ti--Cr ($-7.5$\,kJ/mol), driven by the large electronegativity contrast between early and late refractory elements. Same-group pairs (Ti--Zr, Zr--Hf, Ti--Hf, Nb--Ta, Mo--W) are all within $0.2$\,kJ/mol of zero, reflecting their near-identical electronic structure, while several cross-group pairs are weakly endothermic (Zr--Nb $+4.0$, Hf--Nb $+4.0$, Zr--Ta $+2.8$\,kJ/mol) and V--Ta is strongly so ($+12.4$\,kJ/mol). The Al--refractory interactions are exceptionally strong, with Al--Zr ($-56.0$\,kJ/mol), Al--Hf ($-50.5$\,kJ/mol) and Al--Ti ($-40.6$\,kJ/mol) all exceeding the Al--Ni interaction of the 3$d$ system.

On this common scale the W-containing pairs are revealed to be weak rather than strong. W--Zr ($-9.1$), W--Nb ($-8.4$), W--Ta ($-7.4$) and W--Hf ($-6.4$\,kJ/mol) are comparable to the modest IVB--VIB interactions, and W--Mo ($-0.2$), W--V ($-0.8$) and W--Cr ($+1.0$\,kJ/mol) are essentially athermal. Al--W ($-2.0$\,kJ/mol) is likewise anomalously weak and is the sole exception to the strong Al--refractory affinity, because W has the highest electron density of any element considered ($n_{\mathrm{ws}} = 5.93$) and its positive electron-density mismatch term with Al nearly cancels the charge-transfer term. This weakness is consequential rather than incidental. It means that W, despite its size and its dominant role in setting the melting temperature, is not an enthalpic switch element, and it is the direct reason why the prototypical Senkov alloys NbMoTaW and VNbMoTaW possess no dominant pair and revert to entropic control, as quantified in Section~\ref{sec:dominant}.

The physical origin of these trends follows directly from the variation in Miedema parameter. Group~IVB elements (Ti, Zr, Hf) have low $\phi^*$ (3.45 to 3.80~V) and low $n_{\mathrm{ws}}$ (2.80 to 3.51), while Group~VIB elements (Cr, Mo, W) have higher $\phi^*$ (4.65 to 4.80~V) and higher $n_{\mathrm{ws}}$ (5.18 to 5.93). The resulting charge-transfer and electron-density mismatch energies produce strongly negative interaction enthalpies for IVB--VIB pairs, while within-group similarity cancels both contributions. The Al--early-TM interactions are so strong because Al has both low $\phi^*$ and low $n_{\mathrm{ws}}$, maximizing the contrast with all refractory elements. This has direct structural consequences, since Al preferentially forms Al$_3$Ti, Al$_3$Zr, and Al$_3$Hf intermetallics, which are experimentally observed in Al-containing RHEA systems \cite{whitfield2023rate,wen2021effects}. 

\subsection{Effect of Al content on $\Delta H$}

\begin{figure*}[!htbp]
	\centering
	\includegraphics[width=1\linewidth]{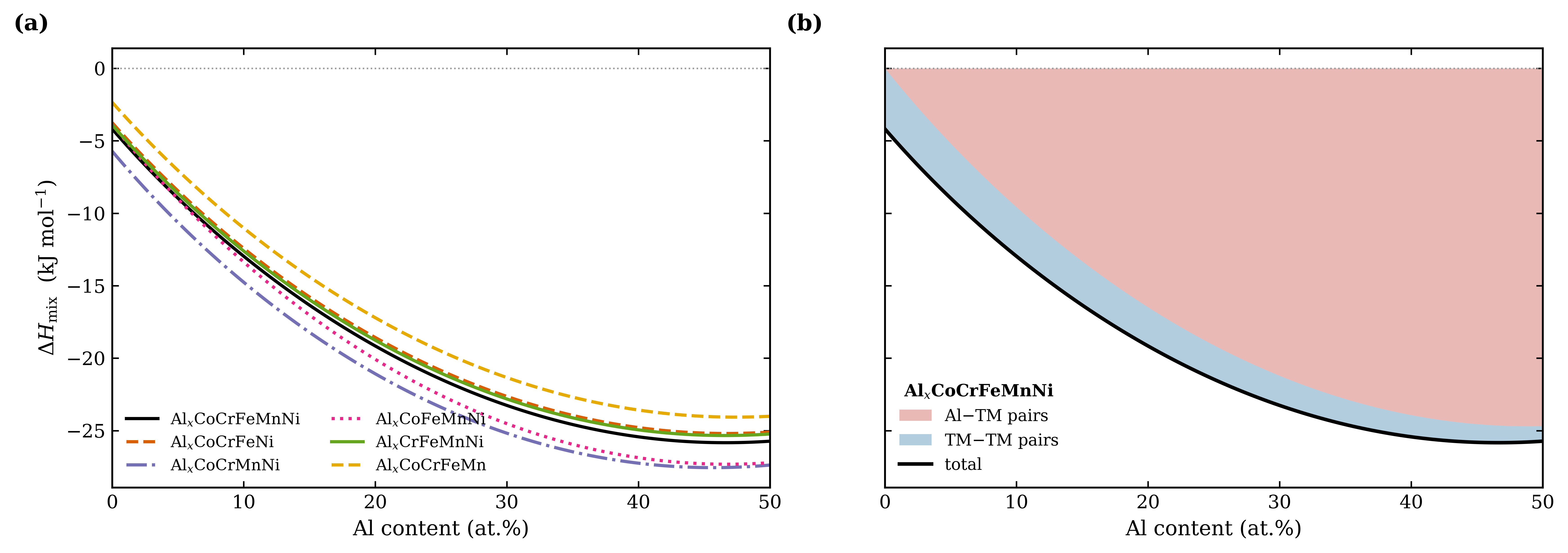}
	\caption{(a) $\Delta H_{\mathrm{mix}}$ against Al content for the senary and the five quinary 3$d$ bases. (b) Stacked decomposition of Al$_x$CoCrFeMnNi into Al--TM and TM--TM pair contributions; the two bands sum exactly to the total.}
	\label{fig:dH_comparison}
\end{figure*}

Figure~\ref{fig:dH_comparison} shows the composition dependence of $\Delta H_{\mathrm{mix}}$ along the Al$_x$ trajectory for the 3$d$ base systems. In Figure~\ref{fig:dH_comparison}(a), all six quinary and quaternary bases exhibit strongly exothermic mixing that becomes progressively more negative with Al addition, reaching between $-25$ and $-29$\,kJ\,mol$^{-1}$ at 50~at.\% Al. The curves remain tightly bunched, spanning only $\sim$4\,kJ\,mol$^{-1}$ across the entire composition range which indicates that the choice of 3$d$ base has a relatively minor influence compared with the Al content itself. The ordering nonetheless carries physical meaning where Cr-free (Al$_x$CoFeMnNi) and Fe-free (Al$_x$CoCrMnNi) bases produce the most negative $\Delta H_{\mathrm{mix}}$. This is because removing one of the weaker-bonding partners increases the relative weight of the strong Al--Ni and Al--Co pairs, whereas the Ni-free base (Al$_x$CoCrFeMn) is the least exothermic, consistent with Al--Ni being the single largest enthalpic contribution in this family.

Figure~\ref{fig:dH_comparison}(b) breaks the total $\Delta H_{\mathrm{mix}}$ of the prototypical Al$_x$CoCrFeMnNi system into Al--TM and TM--TM pair contributions. Two features stand out. First, the TM--TM contribution is small and shrinks toward zero as $x$ increases. This reflects the near-ideal mixing among the 3$d$ transition metals themselves and the geometric reduction in TM--TM pair weighting as Al substitutes for the TM sublattice. Second, the Al--TM contribution grows almost linearly in magnitude and accounts for more than 85\% of $\Delta H_{\mathrm{mix}}$ above 10~at.\% Al, with the total curve tracking the Al--TM envelope closely. These demonstrate Al as the dominant thermodynamic driver in this family where the base-to-base variation reflects only which Al--TM pairs are present and in what proportion, while the overall depth of $\Delta H_{\mathrm{mix}}$ is set almost entirely by the Al--TM interactions.

\subsection{The dominant-element law}
\label{sec:dominant}

\begin{figure*}[!htbp]
	\centering
	\includegraphics[width=1\linewidth]{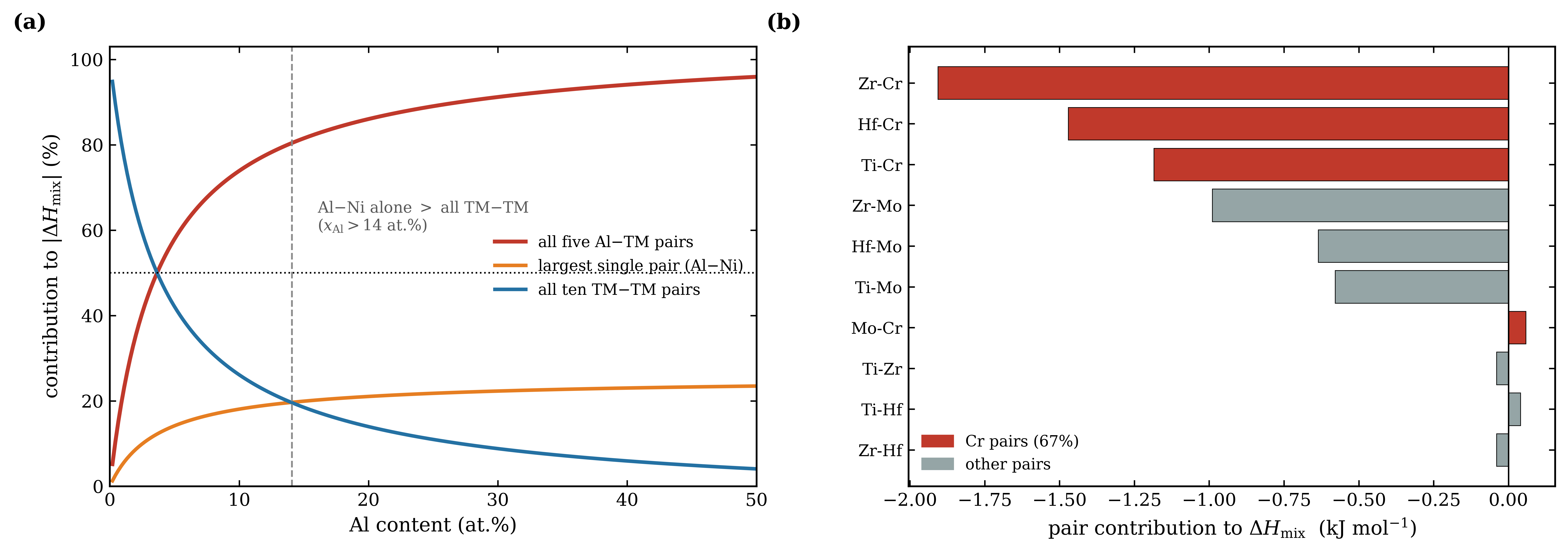}
	\caption{The dominant-element law. (a) In the 3$d$ family Al is the switch, since the five Al--TM pairs exceed 85\% of $|\Delta H_{\mathrm{mix}}|$ above 15\,at.\% Al, and the single Al--Ni pair overtakes all ten TM--TM pairs above $\sim$14\,at.\% Al. (b) In refractory TiZrHfMoCr the switch is Cr, whose three pairs with Ti, Zr and Hf carry 67\% of $\Delta H_{\mathrm{mix}}$.}
	\label{fig:domlaw}
\end{figure*}

The surface-fraction decomposition sharpens the qualitative observation that a few strong pairs dominate into a quantitative and generalizable statement. Figure~\ref{fig:domlaw}(a) tracks the aggregation of the 15 pairwise contributions to the equimolar senary $\Delta H$ along the Al$_x$CoCrFeMnNi trajectory. The five Al--TM pairs collectively exceed 50\% of $|\Delta H|$ by 5\,at.\% Al, 80\% by 15\,at.\%, and $\sim$90\% by 30\,at.\%. Their concentration in one pair is even more pronounced, since the single Al--Ni pair overtakes the sum of all ten TM--TM pairs above $x_{\mathrm{Al}}\approx 14$\,at.\%. The organizing principle is thus not simply that ``a few pairs dominate'' but that phase selection is controlled by the pair sub-block of the single element with the strongest interactions \cite{boakye2026dominant}. In the 3$d$ family this element is Al; in Cr-bearing refractory HEAs it is Cr, whose IVB--VIB contrasts (Zr--Cr, Hf--Cr, Ti--Cr) carry 67\% of $\Delta H$ in TiZrHfMoCr (Fig.~\ref{fig:domlaw}b). The law also predicts its own exceptions. In alloys built entirely from same-group elements with mutually weak interactions (NbMoTaW, VNbMoTaW), no element carries a dominant sub-block, the interactions are near-ideal and partly canceling, and phase selection reverts to entropic control which the framework correctly flags through their large $\Omega$ and small $|\Delta H|$ \cite{boakye2026dominant}.

We emphasize the precise content of the law, because a natural over-reading fails quantitatively. Reconstructing the magnitude of $\Delta H$ from only the two or three largest pairs is not accurate. Across the 27 families studied the mean truncation error of a top-3-pair reconstruction is 28\% (46\% for a top-2 reconstruction). What the top pairs reliably capture is not the numerical value of $\Delta H$ but the identity and sign of the dominant driving force, and hence the direction of phase selection and the ordering of families. Design use of the law is therefore diagnostic (identify the switch element and the sign of its strongest pairs) rather than a shortcut for computing $\Delta H$, for which all pairs involving the switch element (typically 4 to 5 of them) are needed to reach 90\%.

\subsection{Robustness of the enthalpy and of the family separation}
\label{sec:robust}

\begin{figure}[!ht]
	\centering
	\includegraphics[width=1\linewidth]{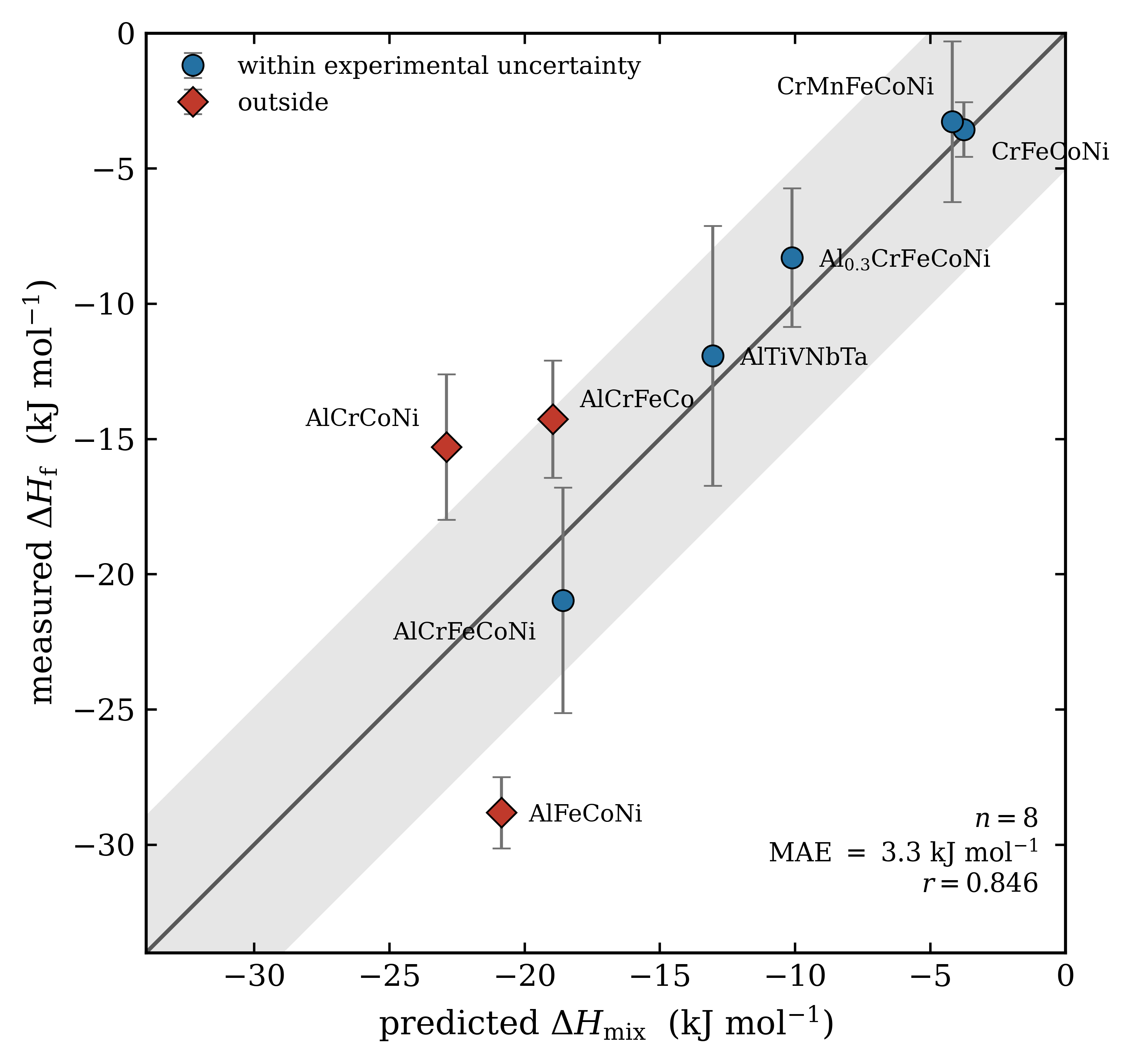}
	\caption{Predicted mixing enthalpies against formation enthalpies measured by high-temperature oxide melt solution calorimetry~\cite{hayun2020enthalpies}, with no adjustable parameters. Error bars are the reported experimental uncertainties and the shaded band spans $\pm 5$\,kJ\,mol$^{-1}$ about parity. Circles lie within one experimental uncertainty and diamonds outside.}
	\label{fig:calorimetry}
\end{figure}

Two checks bound the reliability of the enthalpies underpinning the argument. The first is a comparison against measurement. Calorimetric formation enthalpies for multicomponent alloys are scarce, but Hayun et al.~\cite{hayun2020enthalpies} report them for eight of the systems treated here using high-temperature oxide melt solution calorimetry, providing a benchmark that is independent of the Miedema model. Figure~\ref{fig:calorimetry} compares those measurements with the present calculation, which contains no adjustable parameters. Agreement is close for the alloys on which this work rests, with $-3.75$ against $-3.56 \pm 1.01$\,kJ\,mol$^{-1}$ for CrFeCoNi, $-4.18$ against $-3.27 \pm 2.97$ for CrMnFeCoNi, $-10.11$ against $-8.30 \pm 2.56$ for Al$_{0.3}$CrFeCoNi, $-18.58$ against $-20.97 \pm 4.17$ for AlCrFeCoNi and $-13.04$ against $-11.93 \pm 4.80$ for the refractory AlTiVNbTa. Five of the eight lie within one experimental uncertainty, the mean absolute deviation over the set is $3.3$\,kJ\,mol$^{-1}$ and the correlation is $r = 0.85$.

The three alloys falling outside are the Al-bearing quaternaries AlCrFeCo, AlCrCoNi and AlFeCoNi, where the deviation reaches 2.2 to 6.0 standard deviations and does not carry a single sign, being too negative for the first two and insufficiently negative for the third. A pairwise description of Al in these quaternaries is therefore imprecise, but not biased in a fixed direction, and the deviation does not survive into the quinary alloys that carry the argument. We note further that the binary AlNi and AlCo measurements of the same study refer to ordered B2 compounds, whose formation enthalpies necessarily exceed in magnitude those of the disordered solutions treated here, so they are not a fair comparison and are excluded from the benchmark. None of this bears on the collinearity result, which depends on elemental volumes and valence electron counts rather than on enthalpies at all.

\begin{figure*}[!ht]
  \centering
  \includegraphics[width=1\linewidth]{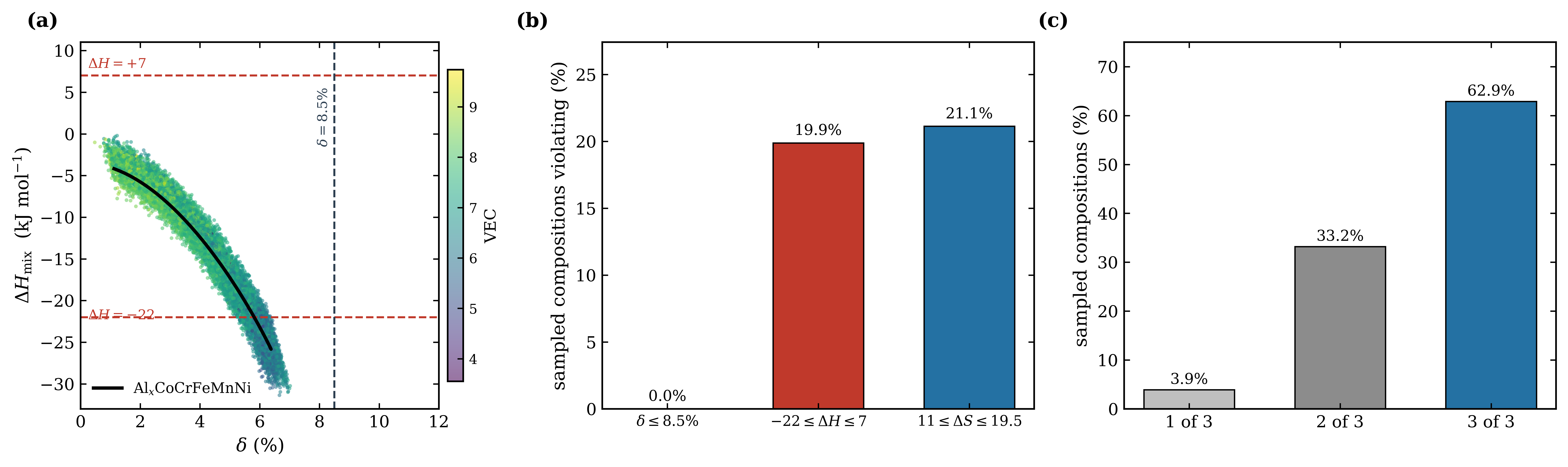}
  \caption{Which Guo--Liu constraint actually binds, over 15{,}000 Dirichlet-sampled 3$d$$+$Al compositions. (a) $\Delta H_{\mathrm{mix}}$--$\delta$ plane colored by VEC, with the criterion boundaries and the Al$_x$CoCrFeMnNi trajectory. (b) Fraction of sampled compositions violating each condition, of which the size criterion is violated by none. (c) Distribution of the three-condition score.}
	\label{fig:multiparameter}
\end{figure*}

The second check concerns which constraint actually binds. Scoring 15{,}000 Dirichlet-sampled 3$d$$+$Al compositions against the three Guo--Liu conditions (Fig.~\ref{fig:multiparameter}), 62.9\% satisfy all three, 33.2\% satisfy two and 3.9\% satisfy one. The distribution of failures is the informative part. Not a single sampled composition violates the size criterion, because Al raises $\delta$ to at most $\sim$6.4\% at 50\,at.\%, well inside the 8.5\% bound, whereas 19.9\% violate the enthalpy window and 21.1\% the entropy window. Within this family the Hume-Rothery size limit is inactive, and solid-solution stability is lost to enthalpy and to the collapse of configurational entropy at the composition extremes rather than to atomic size mismatch. Along the Al$_x$CoCrFeMnNi trajectory the only Guo--Liu condition ever violated is the enthalpy bound, crossed at 26\,at.\% Al. This is a third, independent expression of the same point, since across the accessible composition space atomic size is not the operative constraint on either the driving force or the solid-solution boundary.

\subsection{The two families occupy size-separated islands}

\begin{figure*}[!ht]
	\centering
	\includegraphics[width=1\linewidth]{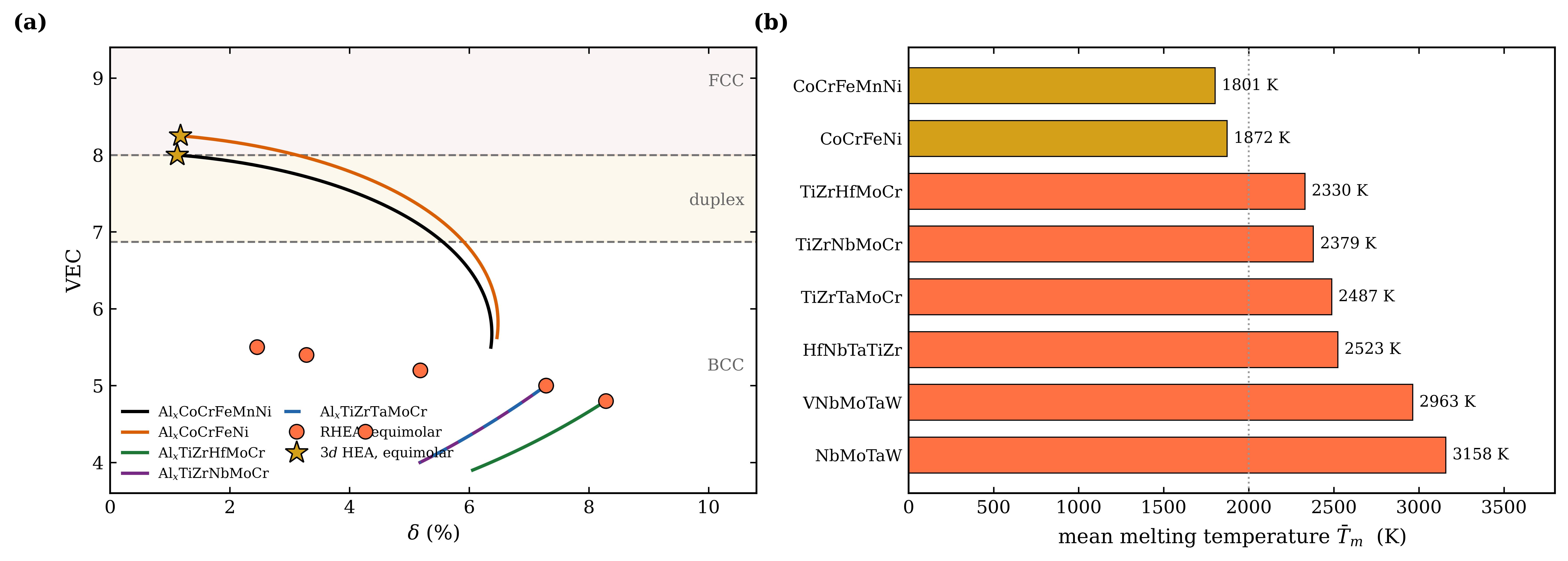}
	\caption{(a) VEC--$\delta$ plane, in which the two families occupy non-overlapping islands. Al$_x$TiZrNbMoCr and Al$_x$TiZrTaMoCr are exactly degenerate here, since Nb and Ta share both an atomic radius and a valence electron count. (b) Mean melting temperature of the same systems at equimolar composition.}
	\label{fig:rhea_vec_Tm}
\end{figure*}

Figure~\ref{fig:rhea_vec_Tm}(a) places both alloy classes on a single VEC--$\delta$ chart. The 3$d$ trajectories sweep from the upper left (VEC $\approx 8$, $\delta \approx 1$\%) toward the lower right as Al is added, while the refractory trajectories lie entirely in the lower-right quadrant, beginning at VEC $\approx 4.5$ to $5.0$ with $\delta \approx 6.5$ to $8$\%. The two families occupy non-overlapping islands, and the gap between them is structural rather than coincidental. The 3$d$ HEAs are built from late transition metals of individual VEC $6$ to $10$, fixing the alloy mean near $8$; the refractory alloys are built from Group~IVB--VIB elements of individual VEC $4$ to $6$, capping the mean near $5.5$. Al addition can lower VEC in either family but cannot raise it, so no compositional path on this plane converts one family into the other.

This separation is the same collinearity of Section~\ref{sec:collinear} seen from a different angle. Both families lie on the master line of Fig.~\ref{fig:collinear}(a); they simply occupy different segments of it, the refractories at low VEC and large $v$, the 3$d$ metals at high VEC and small $v$. That is why the surface correction is large in the Al-bearing 3$d$ alloys, whose compositions straddle a wide span of the line, and negligible in the refractory alloys, whose constituents cluster at one end of it despite their far larger size mismatch. It also explains why BCC-stabilizing strategies developed for 3$d$ Al-HEAs, where Al must overcome a strong FCC bias, do not transfer to refractory alloys in which BCC is structurally guaranteed and Al serves a different role.

Figure~\ref{fig:rhea_vec_Tm}(b) quantifies the second distinction. The systems shown span $\bar{T}_m = 1801$\,K (CoCrFeMnNi) to $3158$\,K (NbMoTaW). Every refractory alloy exceeds $2266$\,K while both 3$d$ alloys fall below $1900$\,K, so even the lowest refractory baseline exceeds the 3$d$ alloys by $\sim$400\,K. Within the present framework this translates directly into stronger entropic stabilization at service temperature.

\subsection{Design implications}

The practical implication of these results is a negative one, and it is nonetheless useful. Atomic size is not an independent lever on phase selection in the alloy families that currently constitute the field. It does not meaningfully move the mixing enthalpy (Section~\ref{sec:decoupling}); its formal effect on the electron count is exactly known but collinear with VEC itself (Sections~\ref{sec:vec}--\ref{sec:collinear}); and it is never the binding constraint on solid-solution stability across the sampled composition space (Section~\ref{sec:robust}). Effort spent constructing size-corrected refinements of VEC within the conventional element set is therefore unlikely to be rewarded, and the more productive route to improved structural prediction lies in enlarging the element set rather than in refining the parameter.

Table~\ref{tab:comparison} summarizes the parametric contrast between the two families at equimolar composition. The most notable entry is that the refractory alloys maintain $\Omega$ values comparable to the 3$d$ alloys despite much larger $\delta$, because their high $\bar{T}_m$ amplifies the $T_m\Delta S$ term, while their elevated $\Delta\chi$ signals a stronger ordering tendency that may limit the single-phase window achievable on annealing.

\begin{table*}[t]
\centering
\caption{Parametric comparison of 3$d$ HEA and RHEA design spaces at equimolar composition.}
\label{tab:comparison}
\small
\begin{tabular}{@{}l c c l@{}}
\toprule
Parameter & 3$d$ HEA & RHEA & Implication \\
\midrule
$\delta$ (\%) & 1.0 to 1.3 & 2.5 to 8.3 & RHEAs span the threshold \\
$\Delta H$ (kJ/mol) & $-3.8$ to $-5.8$ & $-4.2$ to $-7.0$ & Similar magnitude \\
VEC & 7.8 to 8.5 & 4.5 to 5.2 & FCC vs BCC intrinsic \\
$\Psi - \mathrm{VEC}$ & $-0.09$ to $-0.31$ & $-0.02$ to $-0.05$ & Set by covariance, not $\delta$ \\
$\Delta\chi$ & 0.10 to 0.15 & 0.22 to 0.31 & RHEAs order more strongly \\
$\Omega$ & 3.6 to 5.7 & 4.5 to 7.6 & Comparable ($T_m$ compensates) \\
$\bar{T}_m$ (K) & 1520 to 1801 & 2266 to 3158 & RHEA advantage \\
\bottomrule
\end{tabular}
\end{table*}

The row for $\Psi - \mathrm{VEC}$ summarizes the central point. The refractory alloys have the larger size mismatch by a wide margin, yet the smaller size correction to the electron count, because their constituents cluster at one end of the volume--valence line while the Al-bearing 3$d$ alloys straddle a wide span of it. Size mismatch and size sensitivity are therefore distinct quantities, and only the latter is governed by Eq.~(\ref{eq:cov}).

\subsection{Limitations and outlook}

Several limitations should be noted. First, the Miedema model treats only the enthalpic contribution to mixing and assumes random atomic placement. It cannot capture short-range order, which is now well documented in CoCrFeMnNi~\cite{otto2013relative}, and whose neglect likely contributes to the pair-specific overestimation of Al--TM interactions discussed in Section~\ref{sec:robust}.

Second, the collinearity result is a statement about the elements from which HEAs have so far been built, not a law of nature. It is quantitatively strong over the present design set ($r=-0.80$) and weaker over the full tabulation ($r=-0.56$), and it fails outright for the large, electron-rich elements identified in Section~\ref{sec:whyvec}. The scope of the conclusion is therefore the scope of that collinearity, and we have stated explicitly where it ends.

Third, the experimental database, while assembled from 88 independent sources, inherits their heterogeneity of processing route and characterization depth. Phase assignments from as-cast material and from annealed material are not equivalent, and minor phases below the detection limit of laboratory X-ray diffraction are systematically under-reported. We have excluded thin-film and as-deposited entries for this reason, but the residual heterogeneity places a floor on the resolution of any classification comparison, and that floor lies well above the $0.003$ difference in AUC separating VEC from $\Psi$.

Fourth, the present analysis treats only chemical contributions and neglects elastic strain energy, which can be significant for alloys with large $\delta$. This does not affect the collinearity argument, which rests on elemental volumes and valences, but it does bound the accuracy of the enthalpies themselves.

\section{Conclusions}
\label{sec:conclusions}

We set out to explain why the valence electron concentration, a compositional average carrying no geometric information, predicts FCC against BCC stability better than criteria built around atomic size. The mixing enthalpy is almost size-blind, since its surface-fraction correction factorizes exactly into the deviation of the surface fraction from one half multiplied by the dilute-limit asymmetry, and the $2/3$ power holds the first factor below $0.123$. The enthalpy shifts by under 6\% even where constituent volumes differ two-fold, and calorimetric measurements on eight alloys support the calculated values. Any size dependence of phase selection must therefore act on the electron count, where it takes the closed form of the covariance of the atomic surface $V^{2/3}$ with the valence electron count, divided by its mean.

That covariance is not an independent quantity. Atomic surface and valence are strongly anti-correlated across the elements from which high-entropy alloys are built ($r = -0.80$), a consequence of the contraction of the metallic radius with $d$-band filling, so the size-corrected count reduces to an affine rescaling of VEC over 265 characterized alloys and leaves cross-validated discrimination unchanged. VEC succeeds because it is not size-blind, encoding atomic size through this collinearity, and it must fail for elements simultaneously large and electron-rich, none of which appears in our database. Noble-metal alloys are the natural test. Since geometry does not set the enthalpy, chemistry does, through the pair sub-block of a single switch element, Al in the 3$d$ alloys and Cr in  the refractory ones.

\section*{Acknowledgments}
This research was supported by the NSERC Alliance International Catalyst (ALLRP 592696-24), Canada, and the use of a high-performance computing system at the University of Manitoba and the Research Alliances of Canada. 

\section*{CRediT authorship contribution statement}
\textbf{Dennis Boakye}. Writing, review and editing, Writing, original draft, Visualization, Validation, Methodology, Investigation, Formal analysis, Data curation. \textbf{Chuang Deng}. Writing, review and editing, supervision, software, resources, project administration, investigation, Fund Acquisition, conceptualization.

\section*{Declarations}
The authors declare that they have no known competing financial interests or personal relationships that could have influenced the work reported in this paper.

\section*{Data availability}
The experimental phase database compiled for this work is available in the supplementary file.

\section*{Supplementary information}
Supplementary material comprises the annotated experimental phase database (275 entries, 91 sources, with exclusion flags), and the parameter tables.


\bibliography{apssamp}

\end{document}